\documentclass[%
 reprint,
 groupedaddress,
 amsmath,amssymb,
 prl,
floatfix,
superscriptaddress,
]{revtex4-2}

\usepackage{comment}
\usepackage{xcolor}
\usepackage{graphicx}
\usepackage{dcolumn}
\usepackage{bm}
\usepackage{physics}
\usepackage{hyperref}

\usepackage{changepage}

\newcommand{\getJuelichAffiliation}{\affiliation{Institute of Quantum Control (PGI-8), Forschungszentrum Jülich, D-52425 Jülich, Germany}}
\newcommand{\getRegensburgAffiliation}{\affiliation{Institute of Theoretical Physics, University of Regensburg, D-93053 Regensburg, Germany}}
\newcommand{\getKoelnAffiliation}{\affiliation{Institute for Theoretical Physics, University of Cologne, D-50937 Köln, Germany}}

\begin{document}

\title{Comment on ``Beyond-classical computation in quantum simulation''}

\author{Wladislaw Krinitsin}\getJuelichAffiliation\getRegensburgAffiliation
\author{Nikita Alert}\getJuelichAffiliation\getKoelnAffiliation
\author{Matteo Rizzi}\getJuelichAffiliation\getKoelnAffiliation
\author{Markus Schmitt}\getJuelichAffiliation\getRegensburgAffiliation

\begin{abstract}
A recent article [Science 388, 199-204 (2025)] investigates the applicability of numerical methods and a quantum processor unit in simulating a quantum annealing protocol. 
One of the findings indicates that Neural Quantum States---a versatile variational ansatz for the many-body wave function based on artificial neural networks---fail to reach the same accuracy as the quantum processor.
In this comment we revisit these concerns, demonstrating that NQS can provide competitive results in some of the cases when accounting for the Monte-Carlo noise and large autocorrelation times between samples obtained from the final state.
\end{abstract}

\maketitle

In a recent paper, King \textit{et al.}~\cite{king_2025} provided a comprehensive benchmark and comparison of various classical approaches, including tensor networks such as Matrix Product States (MPS), Projected Entangled Pair States (PEPS) as well as Neural Quantum States (NQS) against superconducting quantum processor units (QPU) in performing a quench/annealing dynamics to a spin glass Hamiltonian on several different lattice geometries and for different annealing times.
The quench Hamiltonian takes on the form
\begin{align}
    \mathcal{H}(t) = \Gamma(t/t_a) \mathcal{H}_D + \mathcal{J}(t/t_a) \mathcal{H}_P \\
    \mathcal{H}_D = - \sigma^x, \quad \mathcal{H}_P =  \sum_{i<j} J_{ij}\sigma^z_i \sigma^z_j
\end{align}
with the annealing time $t_a$ and the driving/classical Ising problem Hamiltonian $\mathcal{H}_D$/$\mathcal{H}_P$, respectively. The time dependent parameters start at $\Gamma(0) \gg \mathcal{J}(0)$, corresponding to the paramagnetic phase, and end at $\Gamma(1) \ll \mathcal{J}(1)$ within the spin-glass phase, characterized by random couplings $J_{ij}$ drawn from $[-1,1]$ in steps of $1/128$.

The quality of the time evolution is characterized at $t=0.6\cdot t_a$ by the correlation error 
\begin{equation}
    \varepsilon_c=\sqrt\frac{\sum_{i,j}(c_{ij}-\tilde{c}_{ij})^2}{\sum_{i,j}\tilde{c}_{ij}^2} \label{eq:corr_err}
\end{equation}
where $c_{ij}=\langle \sigma^z_i\sigma^z_j \rangle$ denotes the two-point correlations obtained from the quantum processor or classical method in question and $\tilde c_{ij}$ the ground-truth values obtained from well-converged MPS simulations, with the sums running over all lattice points.
For some of these methods, in particular the QPU and NQS, an exact evaluation of the correlations $c_{ij}$ is either not possible or becomes prohibitively expensive. Therefore, they were estimated by snapshot or Monte Carlo (MC) sampling. In either case the samples are drawn according to the Born distribution of the final state and the sample size was $N_\mathrm{MC}=10^6$.

In the following we focus on the analysis of results obtained using NQS evolved via the time-dependent variational principle (TDVP) and their comparison to the QPU, as performed in~\cite{king_2025} for two-dimensional square lattices up to $L=8$ with periodic boundary conditions in one lattice direction.
The authors concluded that at an annealing time of $t_a=7\,\mathrm{ns}$ and at $L=8$, the NQS fails to achieve the same level of accuracy as the QPU, independent of the number of samples used to estimate each TDVP step, see Fig. S28 in the supplementary material of~\cite{king_2025}.
We aim to investigate and contextualize this claim in terms of the sampling budget. 
Notice that the following analysis will be carried out for the first disorder realization for the couplings $J_{ij}$ on the $L=8$ lattice, as it can be found in the dataset~\cite{dataset_king_2024}.

\begin{figure*}[th!]
    \centering
    \includegraphics[width=1\linewidth]{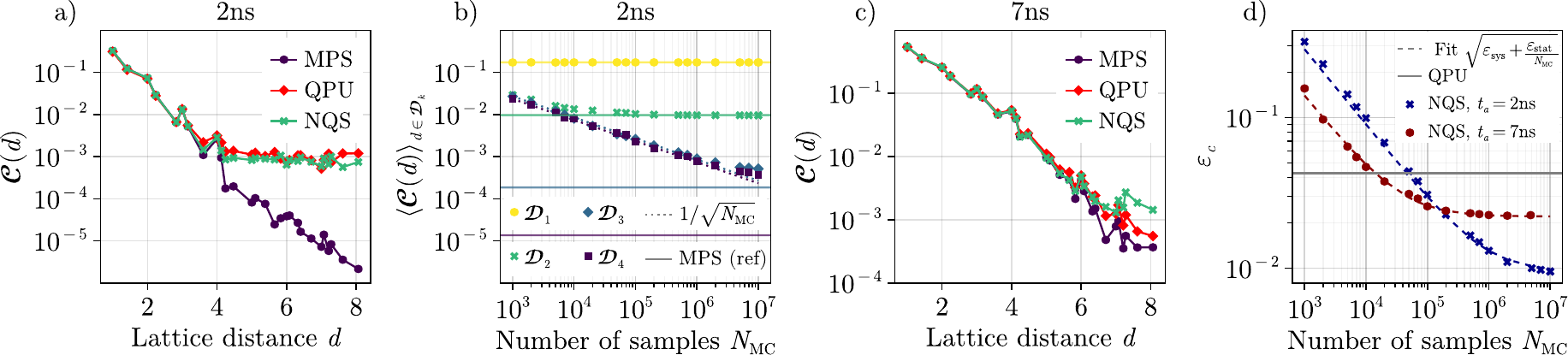}
    \caption{a) Comparison between MPS and NQS of the distance-resolved magnitude of correlations~\eqref{eq:magn_corr} for $t_a=2\, \mathrm{ns}$, using the first disorder realization provided in Ref.~\cite{dataset_king_2024}. The data obtained using NQS plateaus at around $\mathcal{C}(d)\approx 10^{-3}$. b) Behavior of the magnitude of correlations~\eqref{eq:magn_corr}, grouped into four equally-sized distance classes as function of the sampling budget for $t_a=2\, \mathrm{ns}$. At short distances, correlations quickly converge to the MPS reference result, while at larger distances, they are dominated by the MC-estimation noise and drop as $1/\sqrt{N_\mathrm{MC}}$. c) Correlation error~\eqref{eq:corr_err} as a function of the sampling budget for both $t_a=2\, \mathrm{ns}$ and $7\, \mathrm{ns}$. The fitting function~\eqref{eq:corr_error_fit} agrees well with the datapoints for both annealing times.}
    \label{fig}
\end{figure*}

The signs of the correlations $c_{ij}$ depend on the sign structure of the spin-glass groundstate. Therefore, we consider the mean over the absolute values of the two-point correlations correlations, i.e. 
\begin{equation}
    \mathcal{C}(d)=\langle\, |c_{i,j}|\, \rangle_{i,j,|i-j|=d} \label{eq:magn_corr}
\end{equation}
as a measure for the magnitude of the correlations at a certain distance $d$, which, as can be seen in Fig.~\ref{fig}~a) for the MPS ground truth reference result at $t_a=2\, \mathrm{ns}$ taken from~\cite{dataset_king_2024}, drops exponentially with the distance.
Due to the finite velocity of correlation spreading, this is characteristic for the short annealing times of up to $t_a=7\, \mathrm{ns}$: the correlations of the final state remain highly local and the exponential decay means that the correlation function ranges over many orders of magnitude.

We also show the same quantity obtained using an NQS simulation, employing a Restricted Boltzmann machine as the variational ansatz and the same hyperparameters as has been done in~\cite{king_2025}. The correlations were obtained from the time evolved states via a MC estimation with $N_\mathrm{MC}=10^6$ samples.
Interestingly, the data follows the reference data closely up to a lattice distance of around $d\approx3$ or a value of $\mathcal{C}(d)\approx 10^{-3}$, from where it flattens out. 
In order to understand the origin of this behavior, it is instructive to first consider the MC error inherent to an estimation of correlations at larger distances and thus small magnitudes:
\begin{align}
    \sigma_\mathrm{MC}(c_{ij}) = \sqrt{\frac{\mathrm{Var[c_{ij}]}}{N_\mathrm{MC}}} 
    \approx \sqrt{1/N_\mathrm{MC}}, \label{eq:mc_err}
\end{align}
due to the variance becoming
\begin{align}
    \mathrm{Var}[c_{ij}]&=\langle\sigma^z_i\sigma^z_j\sigma^z_i\sigma^z_j\rangle - \langle\sigma^z_i\sigma^z_j\rangle^2 \nonumber \\
    &= 1 - \langle\sigma^z_i\sigma^z_j\rangle^2\approx 1.
    \label{eq:mc_var}
\end{align}
With a sampling budget of $N_\mathrm{MC}=10^6$, the expected MC-noise affecting the estimation of each long-range correlation is of order $\sigma_\mathrm{MC}\approx 10^{-3}$.
This is in line with what we observe in Fig.~\ref{fig}~a): the plateauing at around $\mathcal{C}(d)\approx 10^{-3}$ at larger distances coincides with the size of the MC error. Once the MC error exceeds the absolute value of the correlation function, it drowns the signal.

In order to verify this hypothesis, we take the mean over the distance-resolved magnitudes of the correlations~\eqref{eq:magn_corr}. To that end we define four equally-spaced distance classes $\mathcal{D}_k=[(k-1)\cdot d_\mathrm{max},\ k\cdot d_\mathrm{max}]$ with $d_\mathrm{max}=\sqrt{65}$ being the maximal distance between two points on a two dimensional cylinder of size $(L_x,L_y)=(8,8)$. 
%
Plotting said quantity as a function of the number of samples, see Fig.~\ref{fig}~b), reveals two distinct behaviors: At short distances, i.e., for distance classes $\mathcal{D}_{1/2}$, the correlations quickly converge to the the corresponding quantity obtained from the MPS reference results, denoted by the dotted lines. 
On the other hand at larger distances, i.e. $\mathcal{D}_{3/4}$, the correlations are clearly not converged, following the $1/\sqrt{N_\mathrm{MC}}$ behavior as expected from the MC-estimation error. 
In particular, these quantities have not yet converged to the MPS reference results.
This observation suggests that the estimated values of the comparison metric, i.e. the correlation error~\eqref{eq:corr_err}, do not fully correspond to method-specific limitations, but include significant contributions from statistical noise featuring prominently in the relative error of vanishing correlations at large distances.
Accurate MC estimation of observables is, however, not a limiting factor of the NQS approach.

Nonetheless, it is important to notice that the final state for longer annealing times---including the $t_a=7\mathrm{ns}$ considered in the following---develops a highly peaked Born distribution in the computational basis.
This leads to long autocorrelation times in Markov chain MC sampling that need to be accounted for.
At the same time, the magnitude of correlations $c_{ij}$ increases, such that the relative error at fixed $N_{\mathrm{MC}}$ diminishes, see Fig.~\ref{fig}~c).

In the following we investigate the analytical form of the correlation error~\eqref{eq:corr_err} as a function of $N_\mathrm{MC}$, allowing us to extract the correlation error in the infinite sampling limit.
In the numerator, each squared difference of correlations can be split into two parts: a statistical error contribution of the form $\varepsilon^{(i,j)}_\mathrm{stat}/N_\mathrm{MC}$ and a possible systematic, method-specific error $\varepsilon^{(i,j)}_\mathrm{sys}$.
Notice that for the statistical error contribution, the square root with respect to the number of samples expected from the MC-error cancels with the square from the squared difference in Eq.~\eqref{eq:corr_err}.
By summing up all correlation pairs and absorbing the denominator into the summed error contributions, the decomposition of the correlation error can be brought into the following form:
\begin{equation}
    \varepsilon_c(N_\mathrm{MC}) = \sqrt{\varepsilon_\mathrm{sys}+\frac{\varepsilon_\mathrm{stat}}{N_\mathrm{MC}}}\, \label{eq:corr_error_fit}
\end{equation}
In order to gauge which of the two error contributions, i.e. the systematic or statistical, is more dominant, we evaluate $\varepsilon_c$ for different numbers of samples and both $t_a=2\, \mathrm{ns}$ and $7\, \mathrm{ns}$, and fit a curve parametrized by Eq.~\eqref{eq:corr_error_fit} to the data, see Fig.~\ref{fig}~d).
We find that the fit approximates the data very well, providing a validation to the presented approach. 
In particular, the correlation error we obtain using NQS lies below the one obtained using QPU as well as NQS as presented in Ref~\cite{king_2025}. 
Hence, NQS can provide competitive results for performing annealing protocols, even surpassing the QPU in terms of the correlation error for the considered instance of the disorder realization. 

Clearly, the presented analysis is restricted to the simplest lattice geometry, as well as to $t_a\leq7\mathrm{ns}$; Ref.~\cite{king_2025} on the other hand, shows results obtained using the QPU for more involved geometries and annealing times of up to $t_a=20\mathrm{ns}$. They furthermore consider so-called low-precision inputs, i.e. couplings drawn randomly from a bimodal distribution $J_{i,j}\in\{-1,1\}$, which results in a more rugged energy landscape, for which classical methods, in particular tensor network states optimized using belief propagation, have been shown to exhibit worse correlation errors compared to cases with uniform couplings \cite{nocera2025}. 
While Ref.~\cite{mauron2025} demonstrated favorable performance of a time-dependent variational Monte Carlo approach on larger, diamond-structured lattices, our findings motivate further numerical experiments in order to better understand the strengths and limitations of NQS in a broader scope.

\acknowledgments{The authors acknowledge insightful comments on the results by A.~King and R.~Wiersema. The numerical simulations were performed using the jVMC codebase~\cite{jVMC}.}

\bibliography{ref.bib}%

\end{document}